\documentclass[a4paper]{jpconf}
\usepackage{graphicx,axodraw}
\begin{document}
\title{PAMELA's cosmic positron from decaying LSP \\in SO(10) SUSY
GUT\footnote{Talks presented in PASCOS, SUSY, and COSMO/CosPA in
2010, based on Refs.~\cite{Kyae,DMdecayModel}.}}

\author{Bumseok Kyae}

\address{Department of Physics, Pusan National University, Busan
609-735, Republic of Korea }

\ead{bkyae@pusan.ac.kr}

\begin{abstract}
We propose two viable scenarios explaining the recent observations
on cosmic positron excess. In both scenarios, the present relic
density in the Universe is assumed to be still supported by
thermally produced WIMP or LSP ($\chi$). One of the scenarios is
based on two dark matter (DM) components $(\chi,X)$ scenario, and
the other is on SO(10) SUSY GUT. In the two DM components
scenario, extremely small amount of non-thermally produced
meta-stable DM component [${\cal O}(10^{-10})<n_{X}/n_{\chi}$]
explains the cosmic positron excess. In the SO(10) model,
extremely small R-parity violation for LSP decay to $e^\pm$ is
naturally achieved with a non-zero VEV of $\tilde{\nu}^c$ and a
global symmetry.
\end{abstract}


\section{Introduction}

For a long time thermally produced weakly interacting massive
particles (WIMPs) have been believed to be the most promising dark
matter (DM) candidates. It is because the correct order of the
magnitude of the cross section for explaining the present relic
density of the Universe is naturally possible with WIMPs.
Particularly, the lightest supersymmetric particle (LSP), which is
a well-motivated particle coming from the promising particle
physics model, i.e. the minimal supersymmetric standard model
(MSSM), has attracted much attentions as an excellent example of
WIMP. Actually, the Universe relic density by WIMP or LSP DM is
the traditional scenario, which has been believed so far.

However, recently some astrophysical experimental groups including
PAMELA \cite{PAMELA}, ATIC \cite{ATIC}, and the Fermi-LAT
collaborations \cite{Fermi-LAT} reported the very challenging
observations in cosmic ray: PAMELA observed positron fractions
[$e^+/(e^++e^-)$] exceeding the theoretical expectation
\cite{Moskalenko:1997gh} above 10 GeV upto 100 GeV. On the other
hand, the PAMELA's observations on anti-proton/proton flux ratio
were quite consistent with the theoretical calculation. The ATIC
and Fermi-LAT's observations exhibit excesses of $(e^++e^-)$ flux
in cosmic ray from 100 GeV to 1 TeV. They would result from the
positron flux that keeps rising upto 1 TeV.

Apparently the above observational results are very hard to be
interpreted in view of the conventional MSSM cold DM scenario:
explaining the excess positrons with annihilations of Majorana
fermions such as the LSP needs a too huge boost factor.
Moreover, ATIC and Fermi-LAT's observations seem to require a TeV
scale DM, if they are caused indeed by DM annihilation or decay.
However, TeV scale DM seems to be disfavored by the gamma ray
data, if the excess positron flux should be originated from the
same physics explaining DM creation in the early Universe, i.e.
from DM annihilation \cite{GammaConst}.

Such astrophysical observations seem to destroy our traditional
scenario on DM. In this article, however, it will be pointed out
that such observations do {\it not} necessarily imply the presence
of a new DM theory replacing the traditional one. We will propose
two viable scenarios explaining them, in which the present relic
density in the Universe is assumed to be still supported by
thermally produced WIMP.

\section{Two dark matter components scenario}

Let us consider DM decay scenario to explain the cosmic positron
excess. ``Helicity suppression'' is not valid in DM decay any
longer. Unlike the DM annihilation scenario, DM decay scenario is
relatively free from the gamma ray constraint, since the positron
flux is just linearly proportional to the number density of DM
\cite{strumia}. In the DM decay scenario, however, there are some
serious hurdles to overcome: (1) one is to naturally obtain the
extremely small decay rate of the DM, $\Gamma_{\rm DM}\sim
10^{-26}$ sec.$^{-1}$, and (2) the other is to {\it naturally}
explain the relic density of the DM.

The first hurdle could be somehow resolved by introducing an extra
symmetry, an extra DM component with a TeV scale mass, and grand
unified theory (GUT) scale superheavy particles, which mediate DM
decay into the standard model (SM) charged leptons (and the LSP)
\cite{DMdecayModel}. It is because the required decay rate of
$10^{-26}~{\rm sec.}^{-1}$ can be achieved, if the dominant
operator for decay to $e^+e^-$ is dimension 6 suppressed by
$M^2_{\rm GUT}$, i.e. a four fermion interaction:
\begin{eqnarray}
\Gamma_{\rm DM} \sim \frac{m_{\rm DM}^5}{196\pi^3M^4_{\rm GUT}}
\sim 10^{-26}~{\rm sec.}^{-1} ~,
\end{eqnarray}
where $m_{\rm DM}\sim$ a few TeV. The fact that the GUT scale
particles are involved in the DM decay might be an important hint
supporting GUT \cite{flippedSU5}. However, since the interaction
between the new DM and the SM charged lepton are made extremely
weak by introducing superheavy particles mediating the DM decay,
it is hard to thermally produce the new DM, and so non-thermal
production of DM with a carefully tuned reheating temperature
should be necessarily assumed for the required relic density. One
way to avoid it is to consider a SUSY model with two DM components
$(\chi, X)$, where $\chi$ is just the ordinary WIMP such as the
LSP and $X$ indicates a new DM component
\cite{DMdecayModel,flippedSU5}:

$~~$

\noindent {\bf $\chi$ : main component of DM explaining the relic
density, thermally produced, \\$~~~~~$ absolutely stable.}


\noindent {\bf $X$ : (extremely) small number density,
non-thermally produced, \\ $~~~~~$ meta-stable, decay to $e^+e^-$
explaining PAMELA/Fermi-LAT.}

$~~$

In this class of models, a (global) symmetry should be introduced
to forbid all unwanted dimension 4 and 5 operators contributing to
DM decay. The required number density of $X$ ($\equiv n_X$) turns
out to be just ${\cal O}(10^{-10})<n_{X}/n_{\chi}$. It is possible
only if the suppression of the dimension 6 operator is smaller
than $M_{\rm GUT}$, $10^{12}~{\rm GeV}~<~M_*~<~10^{16}~{\rm GeV}$
\cite{DMdecayModel}, since the positron flux in the case of DM
decay is proportional to $n_X\cdot \Gamma_X$.
The low energy field spectrum in this class of
models~\cite{DMdecayModel} is the same as that of the MSSM except
for the neutral singlet extra DM component.
Moreover, the models of \cite{DMdecayModel} can be embedded in the
flipped SU(5) GUT.

\section{LSP DM scenario: SO(10) SUSY GUT model}

In the second scenario, we suppose again that just the
conventional bino-like LSP is the main component of the DM. Since
the ``bino'' is a WIMP, thermally produced binos could explain
well the relic density of the Universe. Without introducing a new
DM component and interaction, we will attempt to explain the
PAMELA's observation within the framework of the already existing
particle physics model, SO(10) SUSY GUT.

\subsection{SU(5) vs. SU(2)$_{R}$ scales}

In terms of the SM's quantum numbers, the SO(10) generator ($={\bf
45}_G$) is split into the generators of the SM gauge group plus
$\{{\bf (1,1)}_{-1}, {\bf (1,1)}_{1}\}$, ${\bf (1,1)}_0$, $\{{\bf
(3,2)}_{-5/6},{\bf (\overline{3},2)}_{5/6}\}$, and $\{{\bf
(3,2)}_{1/6},{\bf (\overline{3},2)}_{-1/6};~{\bf (3,1)}_{2/3},{\bf
(\overline{3},1)}_{-2/3}\}$.  We will simply write them as
\begin{eqnarray}
 \{E,E^c\}~,~~N~,~~\{Q',Q^{'c}\}~,~~
\{Q,Q^c~;~U,U^c\}~,
\end{eqnarray}
respectively.
The SM gauge group's generators and $\{E,E^c\}$, $N$ compose the
generators of
SU(3)$_c\times$SU(2)$_L\times$SU(2)$_R\times$U(1)$_{B-L}$ ($\equiv
$ LR), where $\{E,E^c\}$ and a linear combination of the SM
hypercharge generator and $N$ ($\equiv N_R$) correspond to the
SU(2)$_R$ generators. The other combination orthogonal to it is
the U(1)$_{B-L}$ generator ($\equiv N_{BL}$). {\it Note that
$\{E,E^c\}$ and $N$ don't carry color charges.} By the VEV of the
adjoint Higgs $\langle {\bf 45}_H\rangle$, the SO(10) gauge
symmetry may break to LR. On the other hand, the SM gauge group's
generators and $\{Q',Q^{'c}\}$ compose the SU(5) generators. The
VEVs of the Higgs in the spinorial representations $\langle {\bf
16}_H\rangle$, $\langle \overline{\bf 16}_H\rangle$ breaks SO(10)
down to SU(5).
Hence, $\langle {\bf 45}_H\rangle$ and $\langle {\bf 16}_H\rangle$
determine the SU(5) and LR breaking scales, respectively.

Non-zero VEVs of both $\langle {\bf 45}_H\rangle$ and $\langle
{\bf 16}_H\rangle$ eventually give the SM gauge symmetry at low
energies.\footnote{Alternatively, one can employ the large
representations, ${\bf 126}_H$, $\overline{\bf 126}_H$, and ${\bf
210}_H$, instead of ${\bf 16}_H$, $\overline{\bf 16}_H$, and ${\bf
45}_H$. ${\bf 126}_H$ and $\overline{\bf 126}_H$ break SO(10) to
SU(5), while ${\bf 210}_H$ breaks SO(10) to
SU(4)$_c\times$SU(2)$_L\times$SU(2)$_R$. In our discussion
throughout this article, ${\bf 16}_H$ ($\overline{\bf 16}_H$) and
${\bf 45}_H$ can be replaced by ${\bf 126}_H$ ($\overline{\bf
126}_H$) and ${\bf 210}_H$, respectively.}
Thus, if $\langle {\bf 45}_H\rangle>\langle {\bf 16}_H\rangle
=\langle \overline{\bf 16}_H\rangle \neq 0$ ($\langle {\bf
16}_H\rangle =\langle \overline{\bf 16}_H\rangle > \langle {\bf
45}_H\rangle \neq 0$), the gauge bosons and gauginos of
$\{Q',Q^{'c}\}$ achieve heavier (lighter) masses than those of
$\{E,E^c\}$ and $N$. The masses of the gauge sectors for
$\{Q,Q^{c}~;~U,U^c\}$ would be given dominantly by the heavier
masses in any cases, since both $\langle {\bf 45}_H\rangle$ and
$\{\langle {\bf 16}_H\rangle, \langle \overline{\bf
16}_H\rangle\}$ contribute to their masses.

\subsection{LSP decay in SO(10)}

To achieve the needed extremely small decay rate of the bino-like
LSP $\chi$, we need extremely small R-parity violation {\it
naturally}. $\chi$ can {\it never} decay, if (1) R-parity is
absolutely preserved {\it and} (2) $\chi$ is really the LSP. We
mildly relax these two conditions such that $\chi$ {\it can} decay
by assuming

{\bf (1)} a non-zero VEV of the superpartner of one family of RH
neutrino, $\tilde{\nu}^c_1$ \\ $~~~~~~~~~$ (i.e. R-parity
violation), {\it or}

{\bf (2)} a mass of $\tilde{\nu}^c_1$ lighter than the $\chi$'s
mass, $m_\chi$ (i.e. $\tilde{\nu}^c_1$ LSP).

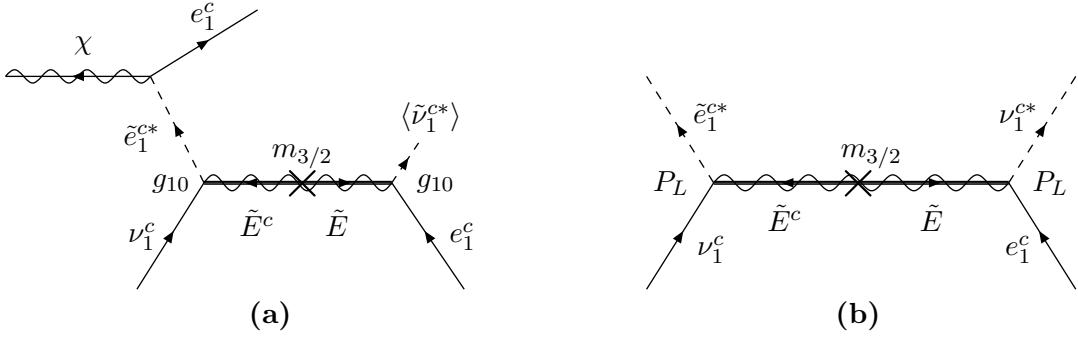
\begin{figure}[t]
\begin{center}
\begin{picture}(400,115)(-10,0)
\ArrowLine(45,90)(-9,90)
\Photon(-9,90)(45,90){2.5}{4}\ArrowLine(45,90)(85,115)
\DashArrowLine(65,50)(45,90){3}
\ArrowLine(100,50)(65,50)\ArrowLine(100,50)(135,50)
\Line(65,50.5)(135,50.5)\Line(65,49.5)(135,49.5)
\Photon(65,50)(135,50){3}{5} \ArrowLine(40,10)(65,50)
\DashArrowLine(135,50)(145,65){3} \ArrowLine(162,10)(135,50)
\Text(102,50)[]{\LARGE $\times$} \Text(102,60)[]{$m_{3/2}$}
\Text(20,102)[]{$\chi$} \Text(150,75)[]{$\langle
\tilde{\nu}^{c*}_1\rangle$} \Text(65,113)[]{$e^c_1$}
\Text(42,67)[]{$\tilde{e}^{c*}_1$} \Text(42,30)[]{$\nu^c_1$}
\Text(162,30)[]{$e^c_1$}
\Text(85,35)[]{$\tilde{E}^c$}\Text(115,35)[]{$\tilde{E}$}
\Text(53,50)[]{$g_{10}$} \Text(152,50)[]{$g_{10}$}
%
\DashArrowLine(255,50)(230,90){3}
\DashArrowLine(365,50)(390,90){3}
%
\ArrowLine(310,50)(255,50) \ArrowLine(310,50)(365,50)
\Line(365,50.5)(255,50.5) \Line(365,49.5)(255,49.5)
\Photon(255,50)(365,50){3}{7} \Text(310,50)[]{\LARGE $\times$}
\ArrowLine(230,10)(255,50) \ArrowLine(390,10)(365,50)
\Text(240,50)[]{$P_L$} \Text(382,50)[]{$P_L$}
\Text(315,60)[]{$m_{3/2}$} \Text(282.5,37)[]{$\tilde{E}^c$}
\Text(337.5,37)[]{$\tilde{E}$} \Text(255,75)[]{$\tilde{e}^{c*}_1$}
\Text(255,25)[]{$\nu^c_1$} \Text(370,75)[]{$\nu^{c*}_1$}
\Text(370,25)[]{$e^c_1$}
\Text(90,0)[]{\bf{(a)}}  \Text(310,0)[]{\bf{(b)}}
\end{picture}
\caption{Dominant diagram of the bino decay (a) and the gauge
interaction between electrically charged superheavy LR gauginos
and the MSSM lepton singlets (b).}\label{fig:gauginoMed}
\end{center}
\end{figure}

\noindent By introducing a global symmetry, one can forbid the
(renormalizable) Yukawa couplings between $\tilde{\nu}^c_1$ and
the MSSM fields. Then, {\it $\tilde{\nu}^c_1$ can interact with
the MSSM fields only through the superheavy gauge fields and
gauginos of SO(10)}, since the (s)RH neutrino $\nu^c_1$
($\tilde{\nu}^c_1$) is a neutral singlet under the SM gauge
symmetry but it is charged under SO(10). It is embedded e.g. in
${\bf 16}$ of SO(10). Consequently, decay of $\chi$ is possible
but extremely suppressed. For instance, refer to the diagram of
Figure \ref{fig:gauginoMed}-(a). We will discuss how this diagram
can be dominant for the $\chi$ decay.


\begin{table}[!h]
\begin{center}
\begin{tabular}{c} \hline
Interactions of the MSSM fields and heavy gauginos \\
\hline\hline
~$\tilde{e}^{c*}_i\nu^c_i\tilde{E}^c$~, ~~
$\tilde{d}^{c*}_iu^c_i\tilde{E}^c$~,~~
$h_u^{+*}\tilde{h}_d^0\tilde{E}^c$~,~~
$h_u^{0*}\tilde{h}_d^-\tilde{E}^c$~
\\
~~$\tilde{\nu}^{c*}_ie^c_i\tilde{E}$~,~~
$\tilde{u}^{c*}_id^c_i\tilde{E}$~,~ ~
$h_d^{0*}\tilde{h}_u^+\tilde{E}$~, ~~
$h_d^{-*}\tilde{h}_u^0\tilde{E}$~
\\ \hline
$\tilde{\nu}^{c*}_i\nu^c_i\tilde{N}$~,
~$\tilde{u}^{c*}_iu^c_i\tilde{N}$~,
~~$h^{+*}_u\tilde{h}^+_u\tilde{N}$~,
~~$h^{0*}_u\tilde{h}^0_u\tilde{N}$
\\
~$\tilde{e}^{c*}_ie^c_i\tilde{N}$ ,\quad
$\tilde{d}^{c*}_id^c_i\tilde{N}$~,
~~$h^{-*}_d\tilde{h}^-_d\tilde{N}$~,
~~$h^{0*}_d\tilde{h}^0_d\tilde{N}$~
\\ \hline\hline
$\tilde{e}^{c*}_iq_i\tilde{Q}^{'c}$~, ~~
$\tilde{d}^{c*}_il_i\tilde{Q}^{'c}$
,~~~$\tilde{q}^{*}_iu^c_i\tilde{Q}^{'c}$~
\\
$\tilde{q}^*_ie^c_i\tilde{Q}'$~, ~~
$\tilde{l}^*_id^c_i\tilde{Q}'$~,~~
$\tilde{u}^{c*}_iq_i\tilde{Q}'$~
\\
\hline
$\tilde{\nu}^{c*}_iq_i\tilde{Q}^{c}$~, ~~
$\tilde{u}^{c*}_il_i\tilde{Q}^{c}$~, ~~
$\tilde{q}^{*}_id^c_i\tilde{Q}^{c}$ ~
\\
$\tilde{q}^*_i\nu^c_i\tilde{Q}$~, ~~
$\tilde{l}^{*}_iu^c_i\tilde{Q}$~, ~~
$\tilde{d}^{c*}_iq_i\tilde{Q}$~
\\
\hline
$\tilde{u}^{c*}_i\nu^c_i\tilde{U}^c$~, ~~
$\tilde{l}^{*}_iq_i\tilde{U}^c$~, ~~
$\tilde{d}^{c*}_ie^c_i\tilde{U}^c$~
\\
$\tilde{\nu}^{c*}_iu^c_i\tilde{U}$~, ~~
$\tilde{q}^{*}_il_i\tilde{U}$~, ~~
$\tilde{e}^{c*}_id^c_i\tilde{U}$~
\\
\hline
\end{tabular}
\end{center}\caption{Gauge interactions between two MSSM fields and a heavy
gaugino in the SO(10) GUT}\label{tab:heavyInt}
\end{table}


\subsection{The conditions for leptonic decay of $\chi$}

Let us consider the interactions of the superheavy
gauginos.\footnote{In this article we don't discuss the cases in
which $\chi$ decays through the mediation of the superheavy {\it
gauge bosons}. However, it turns out that the decay channels of
$\chi$ through the mediation of the superheavy gauge fields are
relatively suppressed, compared to the mediation of the superheavy
gauginos discussed here \cite{Kyae}.} In Table \ref{tab:heavyInt},
we list all the gauge interactions between the superheavy gauginos
of SO(10) and two MSSM fields. They are, of course, the
renormalizable operators. As seen in Table \ref{tab:heavyInt},
$\tilde{\nu}^c_i$ or $\nu^c_i$ couples to the superheavy SO(10)
gauginos, $\{\tilde{E},\tilde{E}^c\}$, $\tilde{N}$,
$\{\tilde{Q},\tilde{Q}^c\}$, and $\{\tilde{U},\tilde{U}^c\}$.

According to PAMELA data \cite{PAMELA}, the branching ratio of the
hadronic DM decay modes should not exceed 10 $\%$. To make the
leptonic interactions, i.e. $\tilde{e}^{c*}_i\nu^c_i\tilde{E}^c$,
$\tilde{\nu}^{c*}_ie^c_i\tilde{E}$, and
$\tilde{\nu}^{c*}_i\nu^c_i\tilde{N}$,
$\tilde{e}^{c*}_ie^c_i\tilde{N}$ dominant over the other
interactions in Table \ref{tab:heavyInt}, we assume that

$~~$

$\bullet$ The LR (or ${\rm B-L}$) breaking scale should be lower
than the SU(5) breaking scale, i.e. $\langle {\bf 16}_H\rangle \ll
\langle {\bf 45}_H\rangle$. Then $M_{Q'}$, $M_{Q}$,
$M_{U}$\footnote{The triplets Higgs contained in ${\bf 10}_h$ can
achieve the mass proportional to $\langle {\bf 45}_H \rangle$ via
${\bf 10}_h{\bf 45}_H{\bf 10}_h$ \cite{minimalSO(10)}.} become
much heavier than $M_E$ and $M_N$, and so most of hadronic decay
modes of $\chi$ can be easily suppressed except those by
$\tilde{E}^c$, $\tilde{E}$, and $\tilde{N}$ in Table
\ref{tab:heavyInt}.

$\bullet$ The slepton $\tilde{e}^c_1$, which composes an SU(2)$_R$
doublet together with $\nu^c_1$, needs to be lighter than the
squarks. Then the decay channels of $\chi$ by
$\tilde{d}^{c*}_iu^c_i\tilde{E}^c$,
$\tilde{u}^{c*}_id^c_i\tilde{E}$, and
$\tilde{u}^{c*}_iu^c_i\tilde{N}$, $\tilde{d}^{c*}_id^c_i\tilde{N}$
become suppressed.  We also require that $\chi$ and
$\tilde{e}^c_1$ are much lighter than the charged MSSM Higgs. So
the leptonic interactions, $\tilde{e}^{c*}_1\nu^c_1\tilde{E}^c$,
$\tilde{\nu}^{c*}_1e^c_1\tilde{E}$, and
$\tilde{\nu}^{c*}_1\nu^c_1\tilde{N}$,
$\tilde{e}^{c*}_1e^c_1\tilde{N}$ can dominate over the others.

$\bullet$ At least one RH neutrino, i.e. the SU(2)$_L$ singlet
neutrino $\nu^c_1$ (and its superpartner $\tilde{\nu}^c_1$) must
be lighter than $\chi$ so that $\chi$ decays to charged leptons.
It is because $\nu^c_i$ is always accompanied by $\tilde{\nu}_i^c$
in the effective operators leading to the leptonic decay of
$\chi$, composed of $\tilde{e}^{c*}_1\nu^c_1\tilde{E}^c$,
$\tilde{\nu}^{c*}_1e^c_1\tilde{E}$, and
$\tilde{\nu}^{c*}_1\nu^c_1\tilde{N}$,
$\tilde{e}^{c*}_1e^c_1\tilde{N}$. If all the sneutrino masses are
heavier than $\chi$, $\tilde{\nu}^c_1$ must develop a VEV for
decay of $\chi$. Once $\nu_1^c$ is light enough, $\tilde{\nu}_1^c$
can achieve a VEV much easily.

$~~$

To be consistent with PAMELA's observations on high energy
galactic positron excess \cite{PAMELA}, the DM mass should be
around 300 -- 400 GeV \cite{decayPAMELA}.\footnote{The $(e^++e^-)$
excess observed by Fermi-LAT could be explained by astrophysical
sources such as nearby pulsars (and/or with the sub-dominant extra
TeV scale DM component \cite{DMdecayModel}). In fact, pulsars can
explain both the PAMELA and Fermi-LAT's data in a suitable
parameter range \cite{profumo}.
However, this does not imply that DM in addition to pulsars can
not be the source of the galactic positrons. In fact, we don't
know yet a complete pulsar model, in which all the free parameters
would be fixed by the fundamental physical constants.

Alternatively, one could assume $m_\chi\approx 3.5$ TeV and the
model is slightly modified such that $\chi$ decays dominantly to
$\mu^\pm,~\nu^c_2$ rather than to $e^\pm,~\nu^c_1$, which is
straightforward, the Fermi-LAT's data as well as the PAMELA's can
be also explained \cite{decayPAMELA}. In this case, however, the
soft SUSY breaking scale should be higher than 3.5 TeV.} Thus, we
simply take the following values;

$~~$

${\bf 1.}$ $\langle {\bf 16}_H\rangle$ $\ll \langle {\bf
45}_H\rangle$. (If $m_\chi < m_{\tilde{\nu}^c_i}$, then $\langle
\tilde{\nu}^c_1\rangle\neq 0$.) We will assume $\langle {\bf
16}_H\rangle \sim 10^{14}$ GeV.

${\bf 2.}$ squarks, charged Higgs, higgsinos and other typical
soft masses are of ${\cal O }(1)$ TeV.

${\bf 3.}$ $m_{\nu^c_1}$ $~<~$ $m_\chi ~\sim ~$ 300 -- 400 GeV
$~<~$ $m_{\tilde{e}^c_1}$ $~\ll~$ ${\cal O}(1)$ TeV.

$~~$

\noindent Consequently, SO(10) is broken first to LR, which would
be the effective gauge symmetry valid below the GUT scale. As seen
from Table \ref{tab:heavyInt}, the gauge interactions by the LR
gauginos (and also gauge fields) preserve the baryon numbers. Even
if the masses of the LR gauginos and gauge fields are relatively
light, their gauge interactions don't give rise to proton decay.

\subsection{Seesaw mechanism}

Although one RH neutrino is light enough, the seesaw mechanism for
obtaining the three extremely light physical neutrinos still may
work. Let us consider the following superpotential;
\begin{eqnarray} \label{lep}
W_{\nu}= y^{(\nu)}_{ij}~l_ih_u\nu^c_j(j\neq 1)
+\frac{1}{2}M_{ij}~\nu^c_i\nu^c_j(i,j\neq 1) ,
\end{eqnarray}
where the Majorana mass term of $\nu^c_i$ could be generated from
the non-renormalizable superpotential $\langle {\bf
\overline{16}}_H\rangle\langle {\bf \overline{16}}_H\rangle {\bf
16}_i{\bf 16}_j/M_P$ ($i,j\neq 1$). Thus, $M_{ij}$ ($\gg\langle
h_u\rangle$) could be determined, if the LR breaking scale by
$\langle {\bf \overline{16}}_H\rangle$ is known. In this
superpotential, we note that one RH neutrino $\nu^c_1$ does not
couple to the MSSM lepton doublets and Higgs. For instance, by
assigning an exotic U(1) R-charge to $\nu_1^c$, one can forbid its
Yukawa couplings to the MSSM superfields. Thus, $\nu^c_1$ would be
decoupled from the other MSSM fields, were it not for the heavy
gauge fields and gauginos of the SO(10) SUSY GUT.

Taking into account only Eq.~(\ref{lep}), one neutrino remains
massless. The two heavy Majorana mass terms of $\nu^c_2$ and
$\nu^c_3$ are sufficient for the other two neutrinos to achieve
extremely small physical masses through the constrained seesaw
mechanism \cite{2RHNseesaw}:
\begin{eqnarray} \label{seesaw}
m_\nu=m_\nu^{T}=-\left(
\begin{array}{ccc}
0 & v_{12} & v_{13}\\
0 & v_{22} & v_{23}\\
0 & v_{32} & v_{33}\\
\end{array} \right)
\left(
\begin{array}{ccc}
0 & 0 & 0\\
0 & M^{-1}_{22} & M^{-1}_{23}\\
0 & M^{-1}_{23} & M^{-1}_{33}\\
\end{array} \right)
\left(
\begin{array}{ccc}
0 & 0 & 0\\
v_{12} & v_{22} & v_{32}\\
v_{13} & v_{23} & v_{33}\\
\end{array} \right) ,
\end{eqnarray}
where $v_{ij}\equiv y^{(\nu)}_{ij}\langle h_u\rangle$, and
$M^{-1}_{ij}$ denotes the inverse matrix of $M_{ij}$. One of the
eigenvalues of $m_\nu$ is zero and the other two are of order
$v^2/M$. Through the diagonalization of the mass matrix in
Eq.~(\ref{seesaw}), the three left-handed neutrinos from the
lepton doublet $l_1$, $l_2$, and $l_3$ can be maximally mixed,
whereas the mixing of the RH neutrinos is only between $\nu^c_2$
and $\nu_3^c$.  A complex phase in $y^{(\nu)}_{ij}$ could make the
leptogenesis possible \cite{2RHNseesaw}.

\subsection{LSP decay rate and the seesaw scale}

Let us consider the following terms in the superpotential;
\begin{eqnarray} \label{nuVEV}
W\supset \frac{1}{M_P}\langle\overline{\bf 16}_H\rangle{\bf
16}_1\Sigma^2+\kappa\Sigma^3,
\end{eqnarray}
where $M_P=2.4\times 10^{18}$ GeV and $\kappa$ is a dimensionless
coupling constant. $\Sigma$ is an SO(10) singlet. We assign e.g.
the U(1) R-charge of $2/3$ to ${\bf 16}_1$ and $\Sigma$, and $0$
to $\overline{\bf 16}_H$. This charge assignment forbids the
renormalizable Yukawa couplings between $\nu^c_1$ and other MSSM
fields carrying integer R-charges.

The scale of $\langle\overline{\bf 16}_H\rangle$ ($=\langle {\bf
16}_H\rangle = M_E/\sqrt{2}g_{10}$) can be determined such that it
is consistent with PAMELA data. The soft mass term of $\Sigma$ and
the A-term of $\kappa\Sigma^3$ in the scalar potential permit a
VEV $\langle\tilde{\Sigma}\rangle\sim m_{3/2}/\kappa$. Then, the
scalar potential generates a linear term of $\tilde{\nu}^c_1$
coming from the A-term corresponding to the first term of
Eq.~(\ref{nuVEV}), $V\supset m_{3/2}^3(\langle \overline{\bf
16}_H\rangle/\kappa^2M_P) \tilde{\nu}^c_1$. The linear term and
the soft mass term of $\tilde{\nu}^c_1$ in the scalar potential
can induce a non-zero VEV of $\tilde{\nu}^c_1$:
\begin{eqnarray}
\langle\tilde{\nu}^c_1\rangle \sim \frac{m_{3/2}}{\kappa^2}\times
\frac{M_E}{M_P} .
\end{eqnarray}
Thus, the decay rate of $\chi$ in Figure \ref{fig:gauginoMed}-(a)
can be estimated:
\begin{eqnarray}
\Gamma_\chi=\frac{\alpha_{10}^2\alpha_Ym_\chi^5}{96M_E^4}
\left(\frac{m_{3/2}\langle
\tilde{\nu}^c_1\rangle}{m_{\tilde{e}^c_1}^2}\right)^2\sim
\frac{\alpha_{10}^2\alpha_Ym_\chi^5}{96M_E^2M_P^2}
\left(\frac{m_{3/2}}{\kappa m_{\tilde{e}^c_1}}\right)^4 \sim
10^{-26}~{\rm sec.}^{-1},
\end{eqnarray}
where $\alpha_{10}$ ($\equiv g_{10}^2/4\pi$) and $\alpha_Y$
[$\equiv g_Y^2/4\pi=(3/5)\times g_1^2/4\pi$, where $g_1$ is the
SO(10) normalized gauge coupling of $g_Y$] are approximately
$1/24$ and $1/100$, respectively. Here, we ignore the RG
correction to $\alpha_{10}$.
300 -- 400 GeV fermionic DM  decaying to $e^\pm$ and a light
neutral particle can fit the PAMELA data \cite{decayPAMELA}. For
$m_\chi\approx $ 300 -- 400 GeV, $(m_{3/2}/\kappa
m_{\tilde{e}^c_1}) \sim 10$, $M_E$ or $\langle {\bf 16}_H\rangle$
is estimated to be of order $10^{14}$ GeV. This is consistent with
the assumption $\langle{\bf 16}_H\rangle\ll\langle{\bf
45}_H\rangle\sim 10^{16}$ GeV.
Therefore, the masses of the other two heavy RH neutrinos, which
do not contribute to the process of Figure
\ref{fig:gauginoMed}-(a), are around $10^{10}$ GeV or smaller in
this case: $W\supset y_{ij}(\langle \overline{\bf
16}_H\rangle\langle \overline{\bf 16}_H\rangle/M_P) {\bf 16}_i{\bf
16}_j (i,j\neq 1)\supset y_{ij}(10^{10}~{\rm GeV})\times
\nu^c_i\nu^c_j (i,j\neq 1)$.

\section{Conclusion}

In this article, we pointed out that the traditional DM scenario
based on thermally produced WIMP do not necessarily conflict with
the cosmic positron excess observed by PAMELA and Fermi-LAT. We
have proposed two viable scenarios based on two DM components and
SO(10) SUSY GUT. Particularly in the SO(10) model, extremely small
R-parity violation for LSP decay could be naturally achieved with
a non-zero VEV of $\tilde{\nu}^c_1$ and a global symmetry.

%

$~~$

$~~$




\end{document}